\documentclass[conference]{IEEEtran}

\makeatletter
\def\ps@headings{%
\def\@oddhead{\mbox{}\scriptsize\rightmark \hfil \thepage}%
\def\@evenhead{\scriptsize\thepage \hfil \leftmark\mbox{}}%
\def\@oddfoot{}%
\def\@evenfoot{}}
\makeatother
\usepackage{graphicx}
\usepackage{tabularx}
\usepackage{times}
\usepackage{amsmath}
\usepackage{amssymb}
\usepackage{cite}
\usepackage{epsfig}
\usepackage{subfigure}
\usepackage[ruled]{algorithm2e}

\begin{document}
\title{On Optimizing Energy Efficiency in Multi-Radio Multi-Channel Wireless Networks}

\author{
\IEEEauthorblockN{
{Lu Liu\IEEEauthorrefmark{1}, Xianghui Cao\IEEEauthorrefmark{1}, Yu Cheng\IEEEauthorrefmark{1} and Li Wang\IEEEauthorrefmark{2}}
\IEEEauthorblockA{
\IEEEauthorrefmark{1}Department of Electrical and Computer Engineering,
Illinois Institute of Technology, USA\\
Email: lliu41@hawk.iit.edu; \{xcao10,cheng\}@iit.edu
}
\IEEEauthorblockA{
\IEEEauthorrefmark{2}School of Electronic Engineering,
Beijing University of Posts and Telecommunications, P.R. China\\
Email: liwang@bupt.edu.cn
}
}
}

\maketitle

\begin{abstract}

Multi-radio multi-channel (MR-MC) networks contribute significant enhancement in the network throughput by exploiting multiple radio interfaces and non-overlapping channels. While throughput optimization is one of the main targets in allocating resource in MR-MC networks, recently, the network energy efficiency is becoming a more and more important concern. Although turning on more radios and exploiting more channels for communication is always beneficial to network capacity, they may not be necessarily desirable from an energy efficiency perspective. The relationship between these two often conflicting objectives has not been well-studied in many existing works. In this paper, we investigate the problem of optimizing energy efficiency under full capacity operation in MR-MC networks and analyze the optimal choices of numbers of radios and channels.  We provide detailed problem formulation and solution procedures. In particular, for homogeneous commodity networks, we derive a theoretical upper bound of the optimal energy efficiency and analyze the conditions under which such optimality can be achieved. Numerical results demonstrate that the achieved optimal energy efficiency is close to the theoretical upper bound.

\end{abstract}


\IEEEpeerreviewmaketitle

\section{Introduction}

The multi-radio multi-channel (MR-MC) networks can significantly augment the capacity of wireless networks. While the throughput of traditional single-radio single-channel (SR-SC) networks are limited by interference on the channel, MR-MC networks provide more transmission opportunities due to multiple radio interfaces equipped on the nodes that can exploit multiple available orthogonal channels. 

The enriched channel and radio resources in MR-MC networks construct a multi-dimensional resource space, where resource allocation can be performed to maximize network throughput \cite{hk12,Zeng,Ji,Cheng12}. For example, the resource allocation scheme proposed in \cite{hk12} provides a joint solution of routing, scheduling and channel-radio assignment which can lead to optimal network throughput. Taking end-to-end delay requirement into account, the authors in \cite{Zeng} formulate the delay-aware throughput optimization problem as a linear program and propose a heuristic algorithm to find feasible scheduling of opportunistic forwarding priorities.

Recently, the energy issues of wireless networks are becoming a significant concern, which draws the attention of researchers to take energy saving as another objective in the resource allocation. In the context of MR-MC networks, there have been a few works on energy-aware design, such as in \cite{GC13,Ava,Tang}. With energy consumption as the objective and throughput requirement as a constraint, an optimization problem is formulated and solved in MR-MC networks in \cite{GC13}. Observing that energy can be saved by turning off idle radios, the authors in \cite{Ava} jointly addressed the channel assignment and routing problem in order to minimize energy consumption without impairing network performance.

Since increasing throughput will probably lead to higher expenditure of energy, throughput maximization and energy minimization usually cannot be achieved simultaneously. To deal with these two conflicting objectives, most works establish their optimization problem by setting one of them as objective while treating the other one as a constraint \cite{GC13,Ju}. In many practical cases, such treatment is reasonable since network users often have certain requirements on the achievable throughput or energy budget. On the other hand, the two objectives can be treated at the same time by looking them in a multi-objective optimization perspective \cite{Jiang,Deva}. The Pareto optimal solutions are derived to provide trade-off between the two objectives \cite{info14}. In order to balance throughput and energy consumption and evaluate the performance in such problems, energy efficiency is usually used as the metric which is defined as the achieved throughput per unit of energy cost \cite{chen,TVT}. 

While the existing works mainly focus on optimizing throughput or energy consumption, the optimality of energy efficiency is not fully investigated. One reason is that the energy efficiency in terms of throughput per unit of energy may not be a convex function. \cite{Rui,Mesh} derive the optimal energy efficiency solution; however the optimization is obtained by fixing either throughput or energy consumption of the system, which essentially returns to the optimization of either throughput or energy. 

In this paper, we target on the theoretic optimal energy efficiency of the network and discuss whether the optima can be achieved in different scenarios.

With the multi-dimensional nature of MR-MC networks, exploiting more radios and channels is promising in increasing the network capacity. However, this is not always the case if the numbers of radios and channels (or C-R configuration) are not properly selected. The authors of \cite{Marina} show that when the network capacity is limited by channel resources, throughput will not increase any more when too many radios are turned on. This will further leads to a decrease of energy efficiency since redundant radios waste a large amount of energy in idle state without contributing to throughput. Conducting transmission on an excessive number of channels may also be inefficient due to the unbalance between radio and channel resources. The possible redundancy and unbalance of resources motivates us to investigate the relationship between C-R configuration and energy efficiency, which is performed by an energy efficiency optimization problem over all the possible C-R configurations of an MR-MC network.

In this paper, we model the energy efficiency of a generic MR-MC network and formulate a resource allocation problem that aims to optimize the network energy efficiency over different channel and radio assignments. Solutions to the optimization problem can provide the maximal energy efficiency and the corresponding C-R configuration. In order to further investigate the optimality of the network energy efficiency, theoretical analysis on achieving the optimum is performed. In particular, under homogeneous commodity flows (i.e., flow demand of each commodity is the same), an explicit expression of optimal energy efficiency is obtained, from which we observe that the network energy efficiency is optimized when each flow takes the shortest path from its source to destination. In practical MR-MC networks, the flow may not be hold within the shortest path. For example, when the numbers of radios and channels are not well balanced, sub-optimal paths may also be used, which degrades the network energy efficiency. It is shown that through selecting the best C-R configuration, the network energy efficiency can be maximized. In sum, this paper has 3-fold contributions as follows.

\begin{enumerate}
  \item We formulate a channel-radio allocation problem aiming to optimize the energy efficiency and determine the appropriate number of radios and channels for general MR-MC networks.
  \item We derive a theoretical upper bound with an explicit expression of energy efficiency in homogeneous commodity networks.
  \item We present the numerical results solved from the optimization problem and compare the results with the theoretical upper bound. We show that the achieved optimal energy efficiency is close to theoretical upper bound.
\end{enumerate}

The remainder of this paper is organized as follows. Section II describes the system model. Section III formulates the energy efficiency optimization problem and Section IV presents some theoretical analysis. Numerical results are presented in Section V. Finally, Section VI gives the conclusion remarks.

\section{Network Model}

In this section, we introduce a tuple-based generic system model for MR-MC networks \cite{hk12,Cheng12}. Consider an MR-MC network, and let $\mathcal{N}$ and $\mathcal{C}$ denote the set of nodes and the set of available channels , respectively. For each node $u$, let $\mathcal{R}_u$ denote the set of radio interfaces equipped on it. At any time instant, one radio interface can be tuned to only one channel. 

To accommodate the multi-dimensional resource space spanned by the radios, channels and transmission links of the MR-MC networks, we represent each resource point as a tuple $p$. A tuple includes the information of the transmitter node $u$ ($u\in \mathcal{N}\backslash \mathcal{D}$), receiver node $v$ ($v\in \mathcal{N}\backslash \mathcal{S}$), the radios used by $u$ and $v$, and the channel they are operating on, where $\mathcal{S}$ and $\mathcal{D}$ are sets of flow sources and destinations, respectively. A tuple can be interpreted as that the real physical link $(u,v)$ can be mapped to $|\mathcal{R}_u|\times |\mathcal{R}_v|\times |\mathcal{C}|$ tuples, where $|\cdot|$ denotes the cardinality of a set. The mapping from single link to multiple tuples also explains the nature of how MR-MC networks enrich the resource space. Considering the available radios and channels, we can obtain all the tuples in MR-MC network, which is denoted as $\mathcal{T}$. Let $w_{p_i}$ be the link capacity (transmission rate) of tuple $p_i$. To perform one bit of data transmission on tuple $p_i$, denote $E^t_{p_i}$ and $E^r_{p_i}$ as the amounts of energy spent by the sender and receiver of $p_i$, respectively.

In the context of tuple, there are both co-channel interference and radio conflict that prevent simultaneous transmissions on two tuples.

\begin{enumerate}
  \item {\bf Co-channel interference}: Two tuples are working on the same channel and at least one node of a tuple is located in the interference range of another tuple's transmitter;
  \item {\bf Radio conflict}: Two tuples share at least one common radio of a node.
\end{enumerate}

Based on this, we can characterize the conflict relationship in MR-MC networks with multi-dimensional conflict graph (MDCG) \cite{hk12}, where each tuple acts as a vertex. An edge is set up between two vertices if the corresponding two tuples conflict with each other in either of the aforementioned way. In MDCG, an independent set (IS) stands for a set of tuples that are free of mutual conflict, and can perform transmissions simultaneously. Denote an IS over MDCG as $I$ and the set of all the ISs as $\mathcal{I}$. At any time instant, only the tuples from the same maximum IS can be activated together; therefore the scheduling of the network can be formed by turning on different ISs alternately.

Consider multiple commodity flows in the network with each flow specified by its source-destination pair $(s,d)\in \Omega$, where $\Omega$ is the set of all source-destination pairs. Denote the flow rate of commodity $(s,d)$ as $f^{(s,d)}$. Since each commodity flow will generate traffic on some of the tuples, let $f_{p_i}^{(s,d)}$ denote the amount of traffic on tuple $p_i$ for commodity $(s,d)$.

\section{Energy Efficiency Optimization}

In this section, we formulate the energy efficiency optimization problem. The problem constraints and energy efficiency model are first presented, followed with the procedure to optimize energy efficiency over different numbers of channels and radios in the network.

\subsection{Problem Constraints} \label{formulation}

In this sub-section, we first present necessary problem constraints. Let $\mathcal{T}_{u\to }$ and $\mathcal{T}_{\to u}$ be the set of tuples that take node $u$ as transmitter and receiver, respectively. To formulate the flow constraints, first notice that for each commodity, the amount of traffic injected into the network from source $s$ should be equal to the amount of traffic collected at the destination node $d$. Further, the incoming traffic at an internal node should be equal to the outgoing traffic. Therefore, $\forall (s,d)\in \Omega$, we can obtain the flow conservation constraints as
\begin{align} \label{eq:c1}
f^{(s,d)}&= \sum\limits_{i:p_i \in \mathcal{T}_{s\to }}^{}f_{p_i}^{(s,d)} = \sum\limits_{j:p_j \in \mathcal{T}_{\to d}}^{}f_{p_j}^{(s,d)} \\
\sum\limits_{i:p_i \in \mathcal{T}_{u\to }}^{}f_{p_i}^{(s,d)} &= \sum\limits_{j:p_j \in \mathcal{T}_{\to u}}^{}f_{p_j}^{(s,d)}, \quad \forall u\in \mathcal{N}\backslash \mathcal{S} \backslash \mathcal{D}
\end{align}

Since we consider a slotted system, the scheduling of ISs can be implemented via protocols such as the time-division multiple access (TDMA) protocol.  We denote the fraction of time allocated to IS $I_m$ in one unit slot as $\alpha_m$, which gives constraint
\begin{equation} \label{eq:c2}
\sum_{m=1}^{|\mathcal{I}|} \alpha_m\le 1, \quad 0\le \alpha_m\le 1,\quad \forall m
\end{equation}

Each tuple can only conduct transmission when it is scheduled on. Hence the traffic allocated on a tuple must be transmitted within its active time. The active time of each tuple is bounded as
\begin{equation} \label{eq:c3}
\frac{\sum\limits_{(s,d)\in \Omega}^{}f_{p_i}^{(s,d)}}{w_{p_i}} \le \sum\limits_{m':p_i\in I_{m'}}^{}\alpha_{m'},\quad \forall p_i\in \mathcal{T}
\end{equation}

Finally, the traffic on each tuple should be non-negative
\begin{equation} \label{eq:c4}
f_{p_i}^{(s,d)}\ge 0 \quad \forall i,\quad \forall (s,d)\in \Omega ,\quad \forall p_i\in \mathcal{T}
\end{equation}

\subsection{Energy Efficiency}

We first consider the energy consumption for data transmission, which is spent by the nodes for transmitting or receiving. The flow from a source node $s$ to its destination $d$ will consume energy along the transmission path $\mathcal{P}^{(s,d)}$ (an array of tuples). If the end-to-end flow rate through the path is $f^{(s,d)}$, then within one time slot, each tuple on the path will consume $f^{(s,d)}(E^t_{p}+E^r_{p})$ amount of energy. The energy consumption on the entire path is the sum of energy consumptions on all the tuples along the path. It is also possible that the flow transmission is performed on multiple paths, then the energy consumption for data transmission of flow $f^{(s,d)}$ can be expressed as
\begin{equation} 
E^{(s,d)}=\sum\limits_{i}^{}\sum\limits_{j:p_j\in \mathcal{P}_i^{(s,d)}}^{} f_{\mathcal{P}_i}^{(s,d)} (E^t_{p_j}+E^r_{p_j})
\end{equation}
where $\mathcal{P}_i^{(s,d)}$ is one of the paths involved in transmission and $f_{\mathcal{P}_i}^{(s,d)}$ is the amount of flow on this path. And for the entire network, the energy consumption for data transmission is
\begin{equation} \label{eq:Energy}
E=\sum\limits_{(s,d)\in \Omega}^{}E^{(s,d)}
\end{equation}

When a radio is not scheduled at some instant, we assume it will remain in some low-power mode such as sleep mode to save energy. Suppose the sleep power is $P_0$, then we can obtain the sleep energy consumption within one time slot:
\begin{equation}
E_0=P_0(\sum\limits_{u:u\in\mathcal{N}}^{}|\mathcal{R}_u|-\sum_{m=1}^{|\mathcal{I}|} 2\alpha_m |I_m|)
\end{equation}
where $|I_m|$ is the number of tuples in this IS.

Define energy efficiency as the achieved throughput per unit of energy cost. Ignoring the amount of energy spent on channel/radio/mode switching, the energy efficiency of the network can be given by
\begin{equation}\label{eq:EE}
EE=\frac{\sum\limits_{(s,d)\in \Omega}^{}f^{(s,d)}} {E+E_0 }=\frac{\sum\limits_{(s,d)\in \Omega}^{}f^{(s,d)}} { \sum\limits_{(s,d)\in \Omega}^{}E^{(s,d)}+E_0 }
\end{equation}

\subsection{Energy Efficiency Optimization}

In this section, we optimize the energy efficiency of a network by turning on the most appropriate numbers of channels and radios. For simplicity, assume each node uses the same number of radios ($|\mathcal{R}|$). Suppose each pair of channel and radio numbers ($|\mathcal{C}|,|\mathcal{R}|$) is a \textit{C-R configuration} of the network. The number of C-R configurations of a network is finite, thus the energy efficiency optimization can be done by first calculating the energy efficiency associated with each C-R configuration and then searching for the maximum among all the configurations.

Under one C-R configuration, define the maximum achievable throughput as the capacity of the network. The network is desired to operate under its full capacity. Therefore for each C-R configuration, we calculate its energy efficiency as when the network achieves its full capacity. To obtain the capacity of a C-R configuration, we solve for the maximum throughput. Suppose the flow demand of commodity $(s,d)$ is $f_0^{(s,d)}$. For some or all of the commodities, only a portion of the demand can be achieved. To avoid flow starvation and ensure fairness in the network, we impose the requirement that the achieved fraction of flow demand for all the commodities should be the same, where this fraction is denoted as $\lambda$. The capacity can be obtained by optimizing the objective of throughput $\sum\limits_{(s,d)\in \Omega}^{}\lambda f_0^{(s,d)}$ with constraints from Eq. (\ref{eq:c1} -- \ref{eq:c4}).

In addition to the network capacity (denoted as $f^*$), the solution also provides the resource allocation to the network including scheduling of the tuples and routing. Note that the solution may not be unique, which means there will be multiple resource allocation solutions leading to the same throughput $f^*$. However the energy consumption of each solution may be different. Among these solutions, we select the one with the least energy consumption and omit the other ones since the latter ones are not energy efficient. The selection process is equivalent to minimize energy consumption with capacity as the throughput constraint. This can be done by formulating another optimization problem whose objective is minimizing the total energy consumption $E$ as in Eq. (\ref{eq:Energy}) and including throughput requirement as a new constraint:
\begin{equation} \label{eq:c5}
\sum\limits_{(s,d)\in \Omega}^{}\lambda f_0^{(s,d)} = f^*
\end{equation}

Finally, for each C-R configuration, the energy efficiency can be obtained according to Eq. (\ref{eq:EE}).
After obtained the energy efficiency associated with each C-R configuration, we can search among the results for the optimal efficiency and configuration. 

\section{Optimality Analysis\label{sec4}}

In this section we derive an explicit expression of optimal energy efficiency in the case of homogeneous commodity networks. The optimum also acts as the theoretical upper bound of the achieved energy efficiency. Based on this, the impact of C-R configuration on energy efficiency is analyzed.

\subsection{Theoretical Upper Bound}

As indicated in Eq. (\ref{eq:Energy}), the energy consumption of network is highly related to the transmission paths selected by the flows. If the transmission (reception) power of each tuple is the same, which is denoted as $E^t$ ($E^r$), then the energy consumption for data transmission on each path can be simplified as $|\mathcal{P}^{(s,d)}|f^{(s,d)}(E^t+E^r)$ where $|\mathcal{P}^{(s,d)}|$ is the length (number of hops) of the path from $s$ to $d$. The shorter the path is, the less energy it will consume. In this case, it can be observed that the energy efficiency for a commodity flow $f^{(s,d)}$ can be optimized when the transmission is performed on the shortest path between $s$ and $d$.

Consider a homogeneous commodity network where all the commodities have the same flow rate demand. Under the fairness requirement as mentioned in Section \ref{formulation}, the actual achieved flow rates of all the commodities are also the same. Then optimal energy efficiency can be achieved by different commodities simultaneously, which is also the optimal energy efficiency for the network as shown in Eq. (\ref{eq:EE1}).
\begin{equation}\label{eq:EE1}
EE^*=\frac{1} { \sum\limits_{(s,d)\in\Omega}^{}\frac{1}{|\Omega|}(E^t+E^r)|\mathcal{P}^{(s,d)}|^* }
\end{equation}
where $|\mathcal{P}^{(s,d)}|^*$ is the length of the shortest path for commodity $(s,d)$. In other words, the optimal energy efficiency can be achieved when every commodity takes the shortest path for transmission.

Notice that we omit $E_0$ in Eq. (\ref{eq:EE1}) since this part of energy consumption is relatively small comparing with transmission energy. As a result, the practically achieved energy efficiency will be lower than the optimal value and thereby Eq. (\ref{eq:EE1}) acts as the theoretical upper bound of energy efficiency for an MR-MC network.

\subsection{Impact of C-R Configuration} \label{analysis}

When optimizing the network throughput, turning on more radios or exploiting more channels are plausible since more resources usually will lead to higher throughput (at least no lower throughput). However, this is not always the case in the view of energy efficiency. Based on the previous discussion, the optimality of energy efficiency depends on whether the flow is kept within the shortest path. If the number of radios $|\mathcal{R}|$ and the number of channels $|\mathcal{C}|$ (C-R configuration) of the network are well selected such that these resources can be fully exploited on the optimal path, then optimal energy efficiency can be achieved. Otherwise, if $|\mathcal{C}|$ and $|\mathcal{R}|$ are not well balanced or there are excess resources in the network, the energy efficiency may be impacted.

When the channel resources are limited but $|\mathcal{R}|$ is large, it may result in redundancy of radios. The unused radios will stay in idle state such as sleeping. Although the energy consumed in sleep state is not a significant part compared to the total energy consumption, this part of unnecessary energy expenditure should still be avoided. No mention that the network may not apply power savings for idle radios, consequently the unused radios will stay in idle sensing state which causes a considerable amount of energy waste. Taking this into consideration, $|\mathcal{R}|$ should be no larger than the practically required value. Given $|\mathcal{C}|$, we can solve for the optimal C-R configuration according to the previous section and determine the proper number of $|\mathcal{R}|$. If the node is equipped with more radios than demand, unnecessary radios can be completely shut down to improve the energy efficiency.

If $|\mathcal{C}|$ is so large that the radios on the optimal path cannot fully utilize the channel resources, the network may tend to exploit other paths whose radios are free, in order to further increase the throughput. However in this case the energy efficiency will not be optimal since longer paths are used in the transmissions. Generally, after the optimal path is fully utilized, if there are still available radio and channel resources in the network, it is possible that non-optimal paths will be involved to fully achieve the network capacity, leading to lower energy efficiency. In this case, to maximize energy efficiency as well as maintain in full-capacity status, $|\mathcal{C}|$ should be reduced such that the channel resources can be fully accommodated by the radios on the optimal path.

In the above discussion (and the rest of the paper, if not specified), the bandwidth of each channel is considered irrelevant to $|\mathcal{C}|$. In some scenarios, the system bandwidth is fixed and uniformly divided by the channels. Then the bandwidth of each channel is inversely proportional to $|\mathcal{C}|$, as in Eq. (\ref{eq:C}).
\begin{equation}\label{eq:C}
w_p\propto \frac{1}{|\mathcal{C}|}
\end{equation}
Since the link capacity is constrained by channel bandwidth, smaller $|\mathcal{C}|$ can provide higher link capacity. In this sense, reducing the number of channels (when $|\mathcal{C}|$ is unnecessarily large) can provide a two-fold contribution to the improvement of energy efficiency.

When the C-R configuration is unbalanced but unchangeable, it is still possible to increase energy efficiency if there is no hard requirement on achieving full capacity. If the throughput requirement is slightly decreased from full capacity, the traffic on non-optimal path will become unnecessary and be removed from resource allocation. As a result, the energy efficiency is improved. In the case that no requirement is imposed to throughput, the network can be set to work at maximal energy efficiency point with highest achievable throughput.

The above analysis will be further explained with numerical results in the next section.

\section{Numerical Results}

In this section, we present numerical results of the energy efficiency under different C-R configurations and demonstrate the relationship between C-R configuration and efficiency optimality. We consider two sample topologies with 25 homogeneous nodes deployed in a $1000m\times 1000m$ area. Suppose the number of radios equipped on all the nodes are the same, which varies from 1 to 4, while the number of channels in the system varies from 1 to 8 (which provides 32 C-R configurations for each topology). The physical link capacity of each tuple is set to 1 rate unit. The communication range and interference range of each node are $250m$ and $500m$, respectively. 3 homogeneous commodity flows with the same flow requirement traverse through the network. The unit energy consumptions are set as $E^t+E^r=1$ and $P_0$ is 1\% of transmission power.

\begin{figure}[htbp]
\vspace{-4mm}
  \centering
        \subfigure[Network capacity of topology-1.]{
            \includegraphics*[width=2.5in]{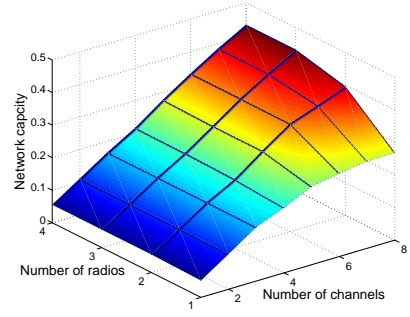}
        \hspace{1mm}
        }
        \subfigure[Energy efficiency of topology-1.]{
            \includegraphics*[width=2.5in]{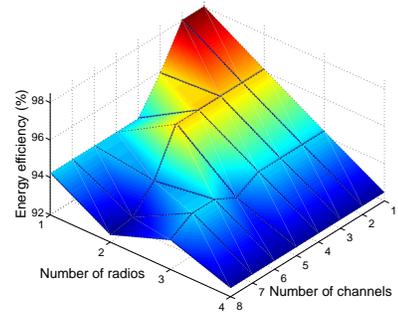}
        }
	\subfigure[Network capacity of topology-2.]{
            \includegraphics*[width=2.5in]{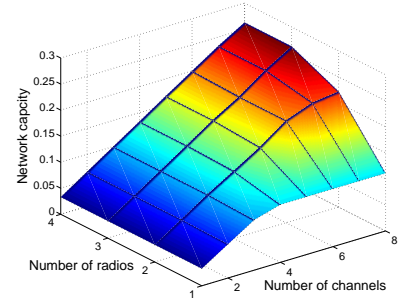}
        }
	\subfigure[Energy efficiency of topology-2.]{
            \includegraphics*[width=2.5in]{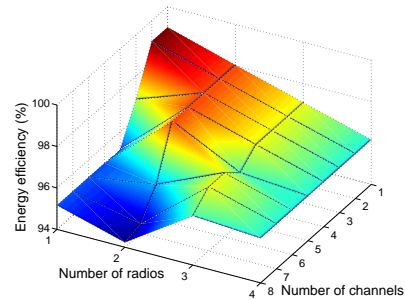}
        }
\vspace{-1mm}
\caption{Network capacity and energy efficiency under different C-R configurations.}\label{fig:1}
\end{figure}

The optimization problem is formulated for each network topology and solved with Cplex \cite{cplex}. In solving the problem, constraint of Eq. (\ref{eq:c4}) is related to the IS scheduling. In order to achieve optimality of solution, it is necessary to explore the entire space of $\mathcal{I}$, which is extremely computationally complex. To tackle this issue, delay column generation (DCG) method is adopted which can avoid rapid growth of the problem scale without affecting the optimality in solutions. The theoretical upper bound of energy efficiency is also obtained based on Eq. (\ref{eq:EE1}) and compared with numerical solutions, where the latter ones are converted to percentage values by dividing the upper bound. Note that since the upper bound cannot be achieved in practice, 96\% or 98\% of the upper bound can be considered as optimal. The network capacity and the corresponding energy efficiency of all the C-R configurations are shown in Fig. \ref{fig:1}

It can be observed that generally higher capacity can be achieved as the number of radios or channels increases. But if the number of channels is limited, more radios will not enhance the capacity. Similarly, if the number of radios is small, too many channels will not significantly increase the capacity since there are not enough radios to utilize channel resource for transmissions.

As for the energy efficiency, optimal efficiency can be achieved when the numbers of radios and channels are well balanced, which appears as a diagonal ridge across the channel-radio space. Especially, energy efficiency is more likely to be optimal when both the numbers of radios and channels are relatively small. When the number of radios is improperly large compared with that of channels, capacity is limited while more energy is consumed by idle radios, resulting in lower energy efficiency. In this case, completely shutting down the redundant radios can save energy and bring energy efficiency back to optimum. When the number of channels is overwhelming, according to the discussion in Section \ref{analysis}, the resource allocation tends to adopt non-optimal transmission paths which will cause the decrease in energy efficiency. Based on the numerical results, we can determine the proper C-R configuration or balance numbers of radios and channels in order to achieve optimal energy efficiency.

\begin{figure}[ht]
  \centering
  \scalebox{0.5}
  {\includegraphics{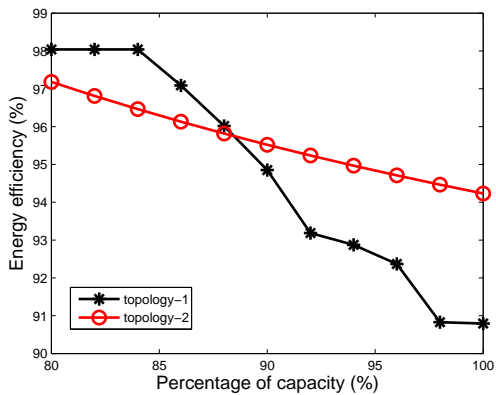}}
  \caption{Relationship between throughput requirement and energy efficiency.}
  \label{fig:2}
\end{figure}

If the C-R configuration is unchangeable, we can still seek for optimal energy efficiency only if there is no need to stick to full-capacity requirement. In this way, energy efficiency can be increased by slightly loosing the throughput constraint, e.g. setting the throughput requirement less than the network capacity in Eq. (\ref{eq:c5}). Fig. \ref{fig:2} shows that under improper number of channels (8 channels and only 2 radios), the adjustment of throughput requirement can contribute to higher energy efficiency.

As shown in Fig. \ref{fig:2}, with full-capacity (100\%), the achieved energy efficiency can be a little bit far from the optimal value since the number of channels is too large comparing with the number of radios. If we keep the C-R configuration but reduce the throughput requirement to 80\% of the network capacity, the achieved optimal energy efficiency is quite close to the upper bound. This way of adjustment provides a trade-off between network throughput and energy efficiency.

\section{Conclusion \label{sec6}}

In this paper, we have formulated the optimization problem to solve the achievable energy efficiency of MR-MC networks under full capacity operation. Theoretical optimal energy efficiency was explicitly expressed for homogeneous commodity networks.  We presented a detailed solution approach to obtain the optimal energy efficiency under full capacity constraint and compared the results with the theoretical upper bound. Numerical results demonstrated the relationship between energy efficiency and C-R configuration. We also showed that, by reducing the throughput requirement, the achieved optimal energy efficiency can be very close to the theoretical upper bound.

\section*{Acknowledgment}
The work of L. Liu, X. Cao and Y. Cheng was supported in part by the NSF under Grant CNS-1320736 and CAREER Award Grant CNS-1053777. The work of L. Wang was supported in part by the National Natural Science Foundation of China (Grant No. 61201150) and the Science Technology Innovation Foundation for Young Teachers in BUPT (Grant No. 2013RC0202).

\bibliographystyle{ieeetr}

\end{document}